\documentclass{aipproc}

\layoutstyle{6x9}

\usepackage{amsmath}
\usepackage{amssymb}
\usepackage{mathbbol}
\usepackage{mathptm}
\usepackage{times}

\begin{document}


\title{Transversity measurements at \textsc{HERMES}}

\author{Markus Diefenthaler (on behalf of the \textsc{HERMES}
collaboration)} 
{address={Physikalisches Institut II,
Friedrich-Alexander-Universit\"at Erlangen-N\"urnberg,
Erwin-Rommel-Stra{\ss}e 1, 91058 Erlangen, Germany},
email={markus.diefenthaler@desy.de}}

\begin{abstract}
Azimuthal single-spin asymmetries (\textsc{SSA}) in semi-inclusive
electroproduction of charged pions in deep-inelastic scattering
(\textsc{DIS}) of positrons on a transversely polarised hydrogen
target are presented. Azimuthal moments for both the Collins and the
Sivers mechanism are extracted. In addition the subleading-twist
contribution due to the transverse spin component from \textsc{SSA} on
a longitudinally polarised hydrogen target is evaluated.
\end{abstract}



\maketitle

Recently the \textsc{HERMES} collaboration published first evidence
for azimuthal single-spin asymmetries (\textsc{SSA}) in the
semi-inclusive production of charged pions on a transversely polarised
target \cite{Airapetian:2004tw}. Significant signals for both the
Collins and Sivers mechanisms were observed in data recorded during
the 2002--2003 running period of the \textsc{HERMES} experiment.
Below we present a preliminary analysis of these data combined with
additional data taken in the years 2003 and 2004. All data was
recorded at a beam energy of \(27.6\text{\texttt{GeV}}\) using a
transversely nuclear-polarised hydrogen-target internal to the
\textsc{HERA} positron storage ring at \textsc{DESY}.
 
At leading twist, the momentum and spin of the quarks inside the
nucleon are described by three parton distribution functions: the
well-known momentum distribution \(q\left(x,Q^{\,2}\right)\), the
known helicity distribution \(\Delta\,q\left(x,Q^{\,2}\right)\)
\cite{Airapetian:2004zf} and the unknown \emph{transversity
distribution} \(\delta\,q\left(x,Q^{\,2}\right)\)
\cite{Ralston:1979ys, Artru:1989zv, Jaffe:1991kp, Cortes:1991ja}. In
the helicity basis, transversity is related to a quark-nucleon forward
scattering amplitude involving helicity flip of both nucleon and quark
(\(N^{\Rightarrow}q^{\leftarrow} \boldsymbol{\rightarrow}
N^{\Leftarrow}q^{\rightarrow}\)). As it is chiral-odd, transversity
cannot be probed in inclusive measurements. At \textsc{HERMES}
transversity in conjunction with the chiral-odd Collins fragmentation
function \cite{Collins:1992kk} is accessible in \textsc{SSA} in
semi-inclusive DIS on a transversely polarised target (\emph{Collins
mechanism}). The Collins fragmentation function describes the
correlation between the transverse polarisation of the struck quark
and the transverse momentum \(\boldsymbol{P}_{\,\text{h}\perp}\) of
the produced hadron. As it is also odd under naive time reversal
(T-odd) it can produce a \emph{\textsc{SSA}}, i.e.~a left-right
asymmetry in the momentum distribution of the produced hadrons in the
directions transverse to the nucleon spin \cite{Burkardt:2003yg}.

The \emph{Sivers mechanism} can also cause a \textsc{SSA}: The T-odd
Sivers distribution function \cite{Sivers:1989cc} describes the
correlation between the transverse polarisation of the nucleon and the
transverse momentum \(\boldsymbol{k}_{\,T}\) of the quarks within. A
non-zero Sivers mechanism provides a non-zero Compton amplitude
involving nucleon helicity flip without quark helicity flip
(\(N^{\Rightarrow}q^{\leftarrow} \boldsymbol{\rightarrow}
N^{\Leftarrow}q^{\leftarrow}\)), which must therefore involve orbital
angular momentum of the quark inside the nucleon
\cite{Burkardt:2003yg, Brodsky:2002cx}.

With a transversely polarised target, the azimuthal angle
\(\phi_{\,\text{S}}\) of the target spin direction in the
``\(\Uparrow\)'' state is observable in addition to the azimuthal
angle \(\phi\) of the detected hadron.  Both azimuthal angles are
defined with respect to the lepton scattering plane.  The additional
degree of freedom \(\phi_{\,\text{S}}\), not available with a
longitudinally polarised target, results in distinctive signatures:
\(\sin{\left(\phi + \phi_{\,\text{S}}\right)}\) for the Collins
mechanisms and \(\sin{\left(\phi - \phi_{\,\text{S}}\right)}\) for the
Sivers mechanism \cite{Boer:1997nt}. Therefore, for all detected
charged pions and for each bin in \(x\), \(z\) or
\(\boldsymbol{P}_{\,\text{h}\perp}\) the cross section asymmetry for
unpolarised beam (U) and transversely polarised target (T) was
determined in the two dimensions \(\phi\) and \(\phi_{\,\text{S}}\):
\begin{equation*}
A_\text{UT}^{\pi^\pm}\left(\phi,\phi_{\,\text{S}}\right) = 
\frac{1}{\left|\,P_z\,\right|}\frac{
N_{\pi^\pm}^{\Uparrow}\left(\phi,\phi_{\,\text{S}}\right) -
N_{\pi^\pm}^{\Downarrow}\left(\phi,\phi_{\,\text{S}}\right)}
{N_{\pi^\pm}^{\Uparrow}\left(\phi,\phi_{\,\text{S}}\right) -
N_{\pi^\pm}^{\Downarrow}\left(\phi,\phi_{\,\text{S}}\right)}.
\end{equation*}
Here \(N_{\pi^\pm}^{\Uparrow\left(\Downarrow\right)}
\left(\phi,\phi_{\,\text{S}}\right)\) represents the semi-inclusive
normalised yield in the target spin state ``\(\Uparrow
\left(\Downarrow\right)\)'', and
\(\left|\,P_z\,\right|=0.754\pm0.050\) denotes the average degree of
the target polarisation.

\begin{figure}
\begin{tabular}{c}
\includegraphics[scale=0.4]{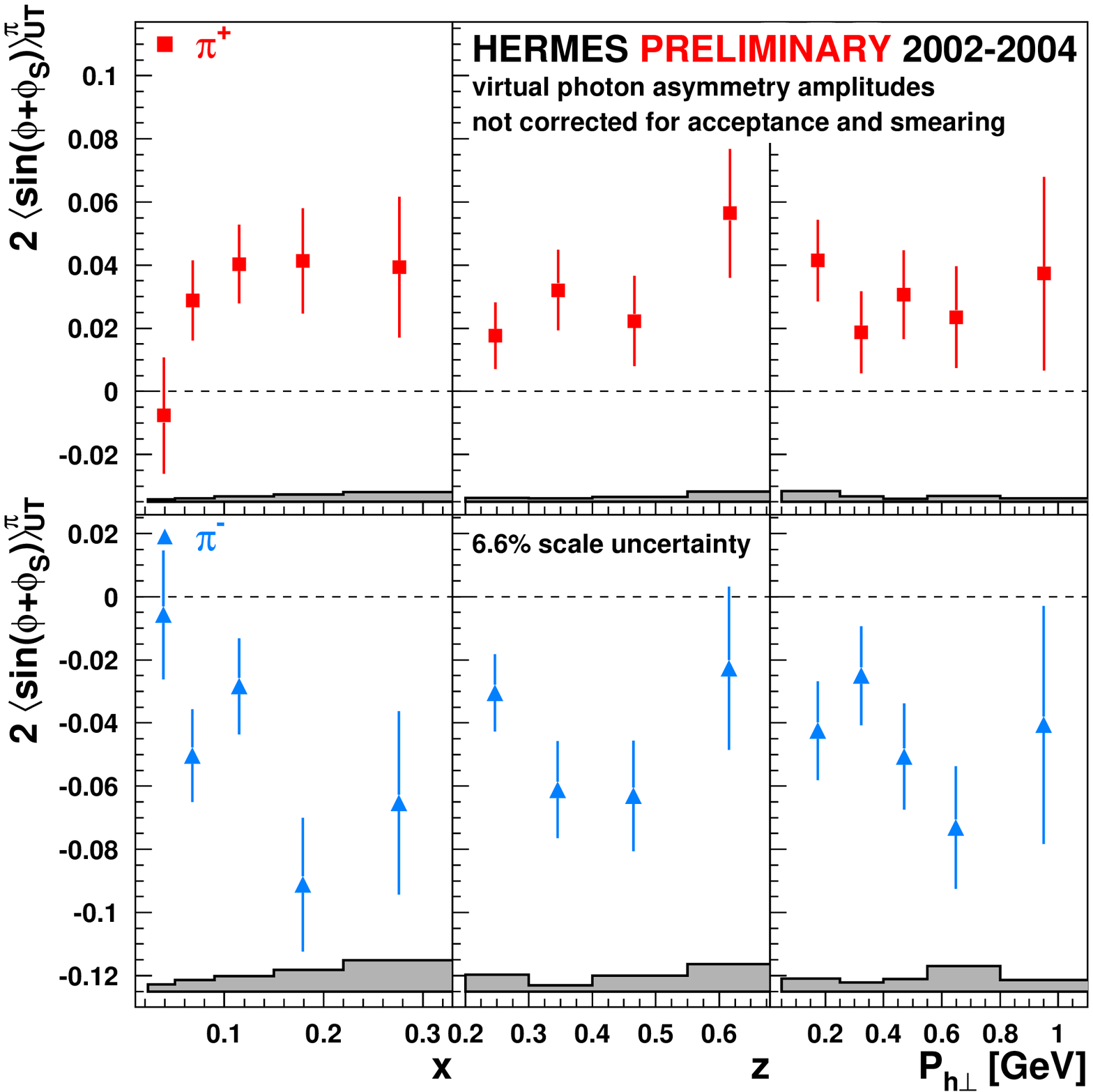} \\
\includegraphics[scale=0.4]{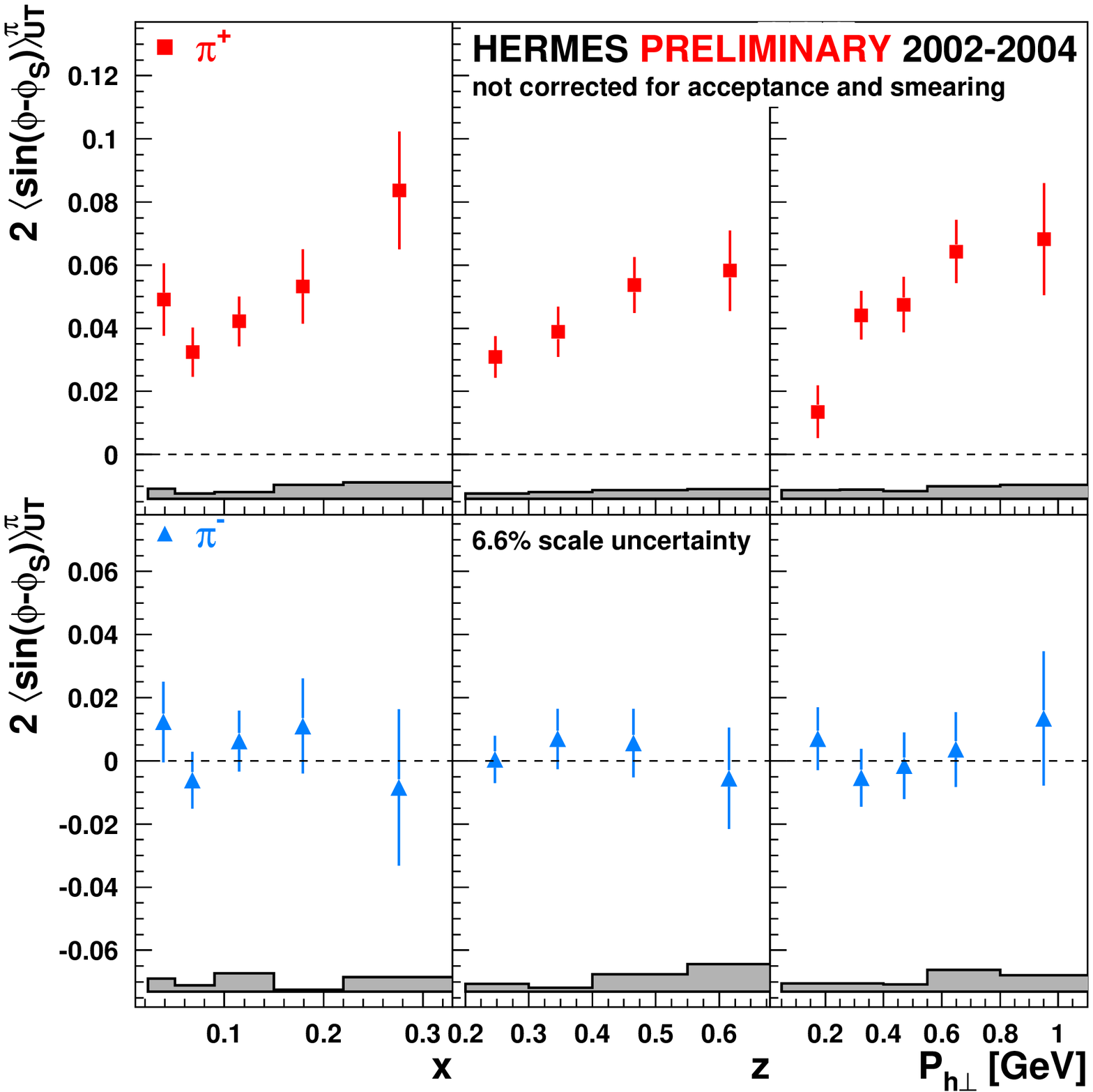}\\
\includegraphics[scale=0.4]{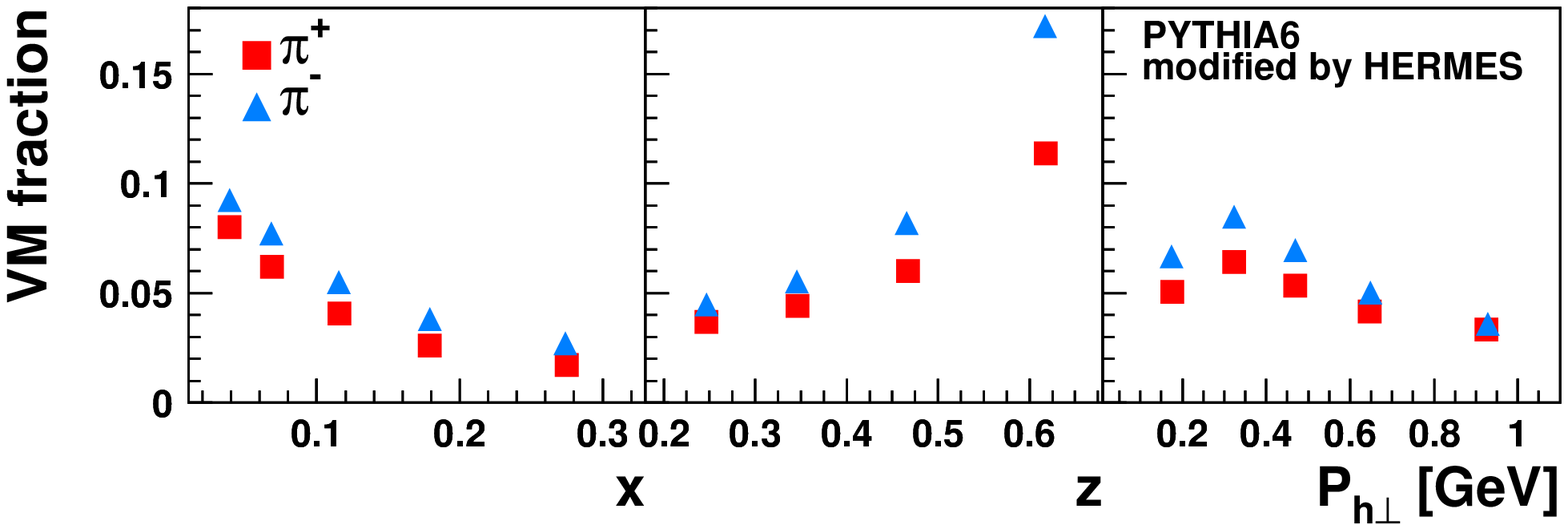}
\end{tabular}
\caption{Collins moments (upper panel) and Sivers moments (middle
panel) for charged pions (as labelled) as a function of \(x\), \(z\)
and \(\boldsymbol{P}_{\,\text{h}\perp}\), multiplied by two to have
the possible range \(\pm1\).  The error bands represent the maximal
systematic uncertainty due to acceptance and detector smearing effects
and due to a possible contribution from the
\(\left<\cos{\phi}\right>_{\text{UU}}\) moment in the spin-independent
cross section. The common overall \(6.6\%\) scaling uncertainty is due
to the target polarisation uncertainty. The lower panel shows the
fraction of charged pions produced in vector meson decay simulated by
\textsc{PHYTHIA6} \cite{Sjostrand:2000wi} (tuned for \textsc{HERMES}
kinematics).}
\label{ssa-moments}
\end{figure}

To avoid cross-contamination, the azimuthal moments for the Collins
mechanism \(\left<\sin{\left(\phi +
\phi_{\,\text{S}}\right)}\right>_{\text{UT}}^{\pi^\pm}\) and the
Sivers mechanism \(\left<\sin{\left(\phi -
\phi_{\,\text{S}}\right)}\right>_{\text{UT}}^{\pi^\pm}\) were
extracted simultaneously. Recent studies showed that the terms for
\(\sin{\phi_{\,\text{S}}}\) and \(\sin{\left(2\phi-
\phi_{\,\text{S}}\right)}\) have to be added in the two-dimensional
fit for the asymmetry (the kinematic factors
\(A\left(\left<\,x\,\right>,\left<\,y\,\right>\right)\) and
\(B\left(\left<\,y\,\right>\right)\) are defined in
\cite{Airapetian:2004tw}):
\begin{eqnarray*}
A_\text{UT}^{\pi^\pm}\left(\phi,\phi_{\,\text{S}}\right) & = &
2\left<\sin{\left(\phi +
\phi_{\,\text{S}}\right)}\right>_{\text{UT}}^{\pi^\pm}
\frac{B\left(\left<\,y\,\right>\right)}{A\left(\left<\,x\,\right>,
\left<\,y\,\right>\right)}
\sin{\left(\phi+\phi_{\,\text{S}}\right)}+\\ &&2\left<\sin{\left(\phi
- \phi_{\,\text{S}}\right)}\right>_{\text{UT}}^{\pi^\pm}
\sin{\left(\phi-\phi_{\,\text{S}}\right)}+\\
&&2\left<\sin{\left(2\phi-\phi_{\,\text{S}}\right)}\right>_{\text{UT}}^{\pi^\pm}
\sin{\left(2\phi-\phi_{\,\text{S}}\right)} +
2\left<\sin{\phi_{\,\text{S}}}\right>_{\text{UT}}^{\pi^\pm}
\sin{\phi_{\,\text{S}}}.
\end{eqnarray*}
The virtual-photon Collins and Sivers moments as a function of \(x\),
\(z\) and \(\boldsymbol{P}_{\,\text{h}\perp}\) are plotted in figure
\ref{ssa-moments} (see caption for the systematic uncertainties). In
addition the simulated fraction of charged pions originating from
diffractive vector meson production and decay is shown, to estimate
the possible contribution from the poorly known asymmetry of this
process.  The average values of the kinematic variables in the
experimental acceptance are \(\left<\,x\,\right>=0.10\),
\(\left<\,y\,\right>=0.53\),
\(\left<\,Q^{\,2}\,\right>=2.43\,\text{GeV}^{\,2}\),
\(\left<\,z\,\right>=0.36\),
\(\left<\,\boldsymbol{P}_{\,\text{h}\perp}\,\right>=0.40\,\text{GeV}\).

This preliminary result is based on nearly five times more statistics
than that in the publication \cite{Airapetian:2004tw} and is
consistent with the published result: The average Collins moment is
positive for \(\pi^+\) and negative for \(\pi^-\). Also, the magnitude
of the \(\pi^-\) moment appears to be not smaller than the one for
\(\pi^+\). The averaged Sivers moment is significantly positive for
\(\pi^+\) and implies a non-vanishing orbital angular momentum of the
quarks inside the nucleon. For \(\pi^-\) the averaged Sivers moment is
consistent with zero.

\begin{figure}
\includegraphics[scale=0.39]{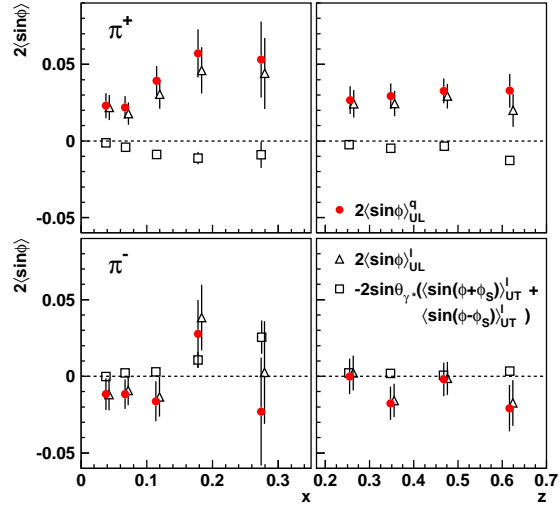}
\caption{The azimuthal moment \(\left<\sin\phi\right>_{\text{UL}}^q\)
(\(\bullet\), multiplied by two) shows the subleading-twist
contribution to the measured asymmetries on a longitudinally polarised
hydrogen target for charged pions as a function of \(x\) and \(z\). In
addition the measured lepton-axis azimuthal moments are plotted
(\(\triangle\) and \(\Box\)). There is an overall systematic error of
\(0.003\). The superscript "q" and "l" distinguishes between
moments with respect to the photon-axis and lepton-axis taking into
account that the measured asymmetries contain contributions from both
transverse and longitudinal polarisation components with respect to
the virtual photon direction \cite{Airapetian:2005jc}.}
\label{ssa-subleadingtwist}. 
\end{figure}

The extracted Collins and Sivers moments allow the evaluation of the
subleading-twist contribution to the previously measured \textsc{SSA}
on a longitudinally polarised hydrogen target \cite{Airapetian:1999tv,
Diehl:2005pc, Airapetian:2005jc}.  These subleading moments
\(\left<\sin\phi\right>_{\text{UL}}^q\) are due to the longitudinal
component of the target spin along the virtual photon direction ("q").
As shown in figure \ref{ssa-subleadingtwist} these moments are almost
the same as the previously published moments
\(\left<\sin\phi\right>_{\text{UL}}^l\), where the longitudinal axis
was defined along the lepton beam momentum ("l").  However, the
maximum contribution of these subleading longitudinal asymmetries to
the leading-twist Collins and Sivers moments in figure
\ref{ssa-moments} is \(0.004\), which is negligible compared to the
statistical uncertainty.

\begin{theacknowledgments}
This work has been supported by the German Bundesministerium f\"ur
Bildung und Forschung (\textsc{BMBF}) (contract nr.~06 ER 125I) and
the European Community-Research Infrastructure Activity under the FP6
''Structuring the European Research Area'' program (HadronPhysics I3,
contract nr.~RII3-CT-2004-506078).
\end{theacknowledgments}

\bibliographystyle{aipproc}

\bibliography{sp-diefenthaler}

\end{document}